\documentclass[superscriptaddress,
  aip,
  pop,
  amsmath,amssymb,
  reprint,
]{revtex4-1}
\usepackage{graphicx,color}
\usepackage{amsmath}
\usepackage{natbib}
\usepackage{epsfig}
\begin{document}

\title{Self-diffusion in single-component Yukawa fluids}

\author{Sergey Khrapak}
\email{Sergey.Khrapak@dlr.de}
\affiliation{Institut f\"ur Materialphysik im Weltraum, Deutsches Zentrum f\"ur Luft- und Raumfahrt (DLR), 82234 We{\ss}ling, Germany}
\affiliation{Aix Marseille University, CNRS, Laboratoire PIIM, 13397 Marseille, France}

\author{Boris Klumov}
\affiliation{Aix Marseille University, CNRS, Laboratoire PIIM, 13397 Marseille, France}
\affiliation{Joint Institute for High Temperatures, Russian Academy of Sciences, 125412 Moscow, Russia}
\affiliation{L. D. Landau Institute for Theoretical Physics, Russian Academy of Sciences, 119334 Moscow, Russia}
\affiliation{Ural Federal University, 620075 Yekaterinburg, Russia}

\author{Lenaic Couedel}
\affiliation{Aix Marseille University, CNRS, Laboratoire PIIM, 13397 Marseille, France}

\date{\today}

\begin{abstract}
It was suggested in the literature that the self-diffusion coefficient of simple fluids can be approximated as a ratio of the squared thermal velocity of the atoms to the  ``fluid Einstein frequency,'' which can thus serve as a rough estimate of the friction (momentum transfer) rate in the dense fluid phase. In this article we test this suggestion using a single-component Yukawa fluid as a reference system. The available simulation data on self-diffusion in Yukawa fluids, complemented  with new data for Yukawa melts (Yukawa fluids near the freezing phase transition), are carefully analyzed. It is shown that although not exact, this earlier suggestion nevertheless provides a very sensible way of normalization of the self-diffusion constant. Additionally, we demonstrate that certain quantitative properties of self-diffusion in Yukawa melts are also shared by systems like one-component plasma and liquid metals at freezing, providing support to an emerging dynamical freezing indicator for simple soft matter systems. The obtained results are also briefly discussed in the context of the theory of momentum transfer in complex (dusty) plasmas.        
\end{abstract}

\maketitle

\section{Introduction}\label{Intro}

De Gennes in his seminal paper on liquid dynamics and inelastic scattering of neutrons~\cite{deGennes1959} related the self-diffusion coefficient in classical atomic liquids to liquid structure, measured in terms  of the radial distribution function (RDF) $g(r)$, and the potential of inter-atomic interactions $\phi(r)$. The relation he put forward is
\begin{equation}\label{Diff_DG}
D = \sqrt{\frac{\pi}{2}}\frac{v_{\rm T}^2}{\Omega_{\rm E}},
\end{equation}    
where $v_{\rm T}=\sqrt{T/m}$ is the thermal velocity, $T$ is the temperature, $m$ is the atomic mass, and the characteristic frequency $\Omega_{\rm E}$ is defined as
\begin{equation}\label{Omega_E}
\Omega_{\rm E}^2=\frac{n}{3m}\int_0^{\infty}d{\bf r}g(r)\Delta \phi(r),
\end{equation}
where $n$ is the liquid number density. This characteristic frequency is called the Einstein frequency, because it represents the oscillation frequency of a single particle in the fixed environment of surrounding particles, characterized by a given RDF.~\cite{HansenBook} de Gennes also pointed out that since the interaction potential was very poorly known for liquids with measured self-diffusion coefficients, a direct comparison was not possible at that time.~\cite{deGennes1959}

The very same characteristic frequency $\Omega_{\rm E}$ was also earlier identified as the friction rate $\xi$ in liquids by Kirkwood, Buff, and Green~\cite{Kirkwood1949} and by Collins and Raffel~\cite{Collins1958} and thus related to the self-diffusion coefficient by means of the Einstein relation, $\xi D\simeq T/m$. This obviously leads to Eq.~(\ref{Diff_DG}) to within the accuracy of a prefactor of order unity. 

These early results from the transport theory of liquids are not expected to be exact, but they nevertheless suggest a sensible way of normalization of the self-diffusion constant,
\begin{equation}\label{Norm}
D_{E}=D\left(\Omega_{\rm E}/v_{\rm T}^2\right),
\end{equation}  
which may be expected to provide a slowly varying quantity in a wide parameter regime. The usefulness of this normalization can be easily tested using present day computational capabilities. In this article we demonstrate this by a detailed investigation of the behavior of $D_{\rm E}$ in one-component Yukawa fluids. Yukawa fluids are particularly suitable for such a test because the Einstein frequency $\Omega_{\rm E}$ is trivially related to an important thermodynamics property, excess internal energy, which is relatively well known in a wide parameter regime. In addition, results related to transport properties of single component Yukawa systems are available in the literature and can be used for extensive tests of earlier predictions. 

In view of existing previous works on transport in Yukawa systems, which also considered self-diffusion in Yukawa fluids, it is important to carefully formulate the motivation for presenting still another article on this topic. Our main purpose here can be divided into the following four objectives: (i) to test quantitatively de Gennes's prediction of Eq.~(\ref{Diff_DG}); (ii) to present new detailed data for the self-diffusion of Yukawa melts; (iii) compare the results for Yukawa melts with available data on diffusion in liquid metals at the melting temperature; and (iv) provide simple explanation on the observed quantitative similarity, which is likely to work for other related systems. To the best of our knowledge, these issues have not been addressed in previous studies.         
 
The article is organized as follows. In Sec.~\ref{Formul}, we provide basic information on the physics of single component Yukawa systems. In Sec.~\ref{Methods}, we outline the procedure of numerical simulations performed in this work. In Sec.~\ref{Results} we summarize our main results and discuss their implications. This is followed by our conclusion in Sec.~\ref{Concl}.   

\section{Formulation}\label{Formul}
 
Yukawa systems are characterized by the pairwise repulsive interaction potential between charged point-like particles of the form
\begin{equation}\label{Yukawa}
\phi(r)=(Q^2/r)\exp(-r/\lambda),
\end{equation}
where $Q$ is the particle charge, $\lambda$ is the screening length, and $r$ is the distance between two particles. The screening normally comes from the redistribution of electrical charges (electron and ions) in a plasma surrounding the particle. Fixed neutralizing background corresponds to the absence of screening and the pure Coulomb interaction potential, the limit known as the one-component plasma (OCP) model. Yukawa potential is considered as a reasonable starting point to model interactions in three-dimensional isotropic complex (dusty) plasmas and colloidal dispersions,~\cite{FortovUFN,FortovPR,IvlevBook,KlumovUFN2010} although it is also well recognized that considerable deviations can occur, especially at long interparticle separations.~\cite{TsytovichUFN1997,KhrapakPRE2001,DeAngelisPoP2006,KhrapakPRL2008,KhrapakCPP2009,ChaudhuriSM2011,LampePoP2015,Fingerprints}

The phase state of Yukawa systems is conventionally described  by the two dimensionless parameters,~\cite{HamaguchiPRE1997} which are the coupling parameter $\Gamma=Q^2/aT$, and the screening parameter $\kappa=a/\lambda$, where $a=(4\pi n/3)^{-1/3}$ is the Wigner-Seitz radius. The screening parameter $\kappa$ determines the softness of the interparticle interaction. It can be varied from the very soft long-ranged Coulomb interaction at $\kappa=0$ to the hard-sphere interaction limit at $\kappa\rightarrow \infty$. In the context of complex plasmas and colloidal suspensions low- and moderate-$\kappa$ regime is of particular interest and this regime will be addressed below.  

Yukawa systems are conventionally called ``strongly coupled'' when the coupling parameter $\Gamma$ is large, that is when the potential interaction energy stored in the system dominates over the kinetic (thermal) energy. In this case the so-called Yukawa fluid regime is realized. When $\Gamma$ increases further (i.e. temperature decreases), Yukawa fluids can crystallize into the body-centered-cubic (bcc) or face-centered-cubic (fcc) lattice (bcc lattice is thermodynamically favorable at week screening). The critical coupling values at which the fluid-solid phase transition occurs are usually denoted $\Gamma_{\rm m}$, where the subscript ``m'' refers to melting. This critical coupling strongly depends on the strength of screening. Accurate values $\Gamma_{\rm m}(\kappa)$ have been tabulated~\cite{HamaguchiJCP1996,HamaguchiPRE1997} and fitted by simple expressions.~\cite{VaulinaJETP2000,VaulinaPRE2002,KhrapakPRL2009} If Yukawa fluid is quickly cooled down to even lower temperatures, a Yukawa glass may form. The glass-transition line appears to be almost parallel to the melting line in the extended region of the phase diagram.~\cite{SciortinoPRL2004,YazdiPRE2014}        

For the Yukawa potential the Einstein frequency is trivially related to the internal excess energy, which is relatively well known analytically.  This relation is the consequence of the identity $\Delta \phi (r)=\phi(r)/\lambda^2$ and can be expressed as ~\cite{RobbinsJCP1988,KhrapakPoP2014}
\begin{displaymath}
\Omega_{\rm E}^2 = \frac{2}{9}\frac{\kappa^2}{\Gamma}\omega_{\rm p}^2u_{\rm ex} = \frac{2}{3}\frac{\kappa^2}{a^2}v_T^2u_{\rm ex}, 
\end{displaymath}
where  $\omega_{\rm p}=\sqrt{4\pi Q^2 n/m}$ is the plasma frequency and $u_{\rm ex}$ is the ecxess energy per particle in units of temperature. Very accurate analytical expressions for the quantity $u_{\rm ex}(\kappa,\Gamma)$ are available in the literature.~\cite{KhrapakPRE2015,KhrapakPPCF2015,ToliasPoP2015,KhrapakJCP2015} Particularly simple practical expressions are based on the Rosenfeld--Tarazona scaling,~\cite{RT,RosenfeldPRE2000} which states that for a wide class of soft repulsive potentials the thermal component of the internal energy exhibits a quasi-universal scaling on approaching the fluid-solid phase transition. The expression for $u_{\rm ex}$ used in this work is that from Ref.~\onlinecite{KhrapakPRE2015}.
 
Self-diffusion in one-component Yukawa fluids has been extensively studied using molecular dynamics (MD) simulations by several authors. In particular, Ohta and Hamaguchi tabulated  the values of the reduced quantity $D/a^2\omega_{\rm E}$ in a wide range of $\kappa$ and $\Gamma$ corresponding to the fluid regime.~\cite{OhtaPoP2000}  The characteristic Einstein  frequency $\omega_{\rm E}$ corresponds to an ideal crystalline lattice and is thus different from that used in this paper [Eq.~(\ref{Omega_E})]. Daligault investigated the diffusion properties of one-component
plasmas and binary ionic mixtures from the weakly to the strongly coupled regimes.~\cite{DaligaultPRL2012} He then presented a practical interpolation formula for the self-diffusion coefficient in Yukawa one-component plasmas valid for a wide range of inverse screening lengths and over the entire fluid region.~\cite{DaligaultPRE2012}
This  practical formula operates with the reduced quantity $D/a^2\omega_{\rm p}$. Rosenfeld used one more normalization $D_{\rm R}=D n^{1/3}/ v_{T}$  to demonstrate that $D_{\rm R}$ (and also similarly reduced shear viscosity) exhibits quasi-universal behavior as a function of reduced excess entropy for various simple model systems,~\cite{RosenfeldPRA1977} including Yukawa fluids.~\cite{RosenfeldPRE2000} The usefulness of the ``fluid Einstein frequency'' $\Omega_{\rm E}$ to normalize self-diffusion data was not explored previously.  

\section{Methods}\label{Methods}

In this paper we use the data for the self-diffusion coefficient in Yukawa fluids tabulated by Ohta and Hamaguchi in a wide range of $\kappa$ and $\Gamma$ values corresponding to the fluid state.~\cite{OhtaPoP2000} In addition, we performed complementary molecular dynamics (MD) simulations for Yukawa fluids at the melting temperature (Yukawa melts). The melting temperatures or, equivalently, coupling parameters $\Gamma_{\rm m}$ are those tabulated for different $\kappa$ in Ref.~\onlinecite{HamaguchiPRE1997}. Our MD simulations were performed on graphics processing unit (NVIDIA GeForce GTX 960) using the HOOMD-blue software.~\cite{HOOMD1,HOOMD2,HOOMD3} We used $N = 50653$ particles interacting according to the pairwise potential (\ref{Yukawa}) and placed in a cubic box with periodic boundary conditions. There was no need to introduce hard cores since the potential is repulsive and diverges at short distances. At long distances the potential was cut off at $L_{\mathrm{cut}}\simeq 14.5 \lambda$. The numerical time step was set to
$\Delta t=5\times 10^{-3}\sqrt{m a^3/Q^2}$. Simulations were performed in the canonical ensemble ($NVT$) with the Langevin thermostat at a temperature corresponding to the desired target coupling parameter $\Gamma$. The dissipation rate used was small enough to ensure that the system dynamics is not affected by dissipation (i.e. we are dealing with the Newtonian dynamics limit, opposite to the over-damped Brownian limit).~\cite{VaulinaPRE2002,PondSM2011,KhrapakPoP2012}
Equilibrium of the system was first achieved by running the simulation for at least two million time steps (depending on the value of $\kappa$ we also used longer equilibration runs). Then, the particle positions and velocities were saved for $300\Omega_{\rm E}^{-1}$  with a resolution of $(1/20)\Omega_{\rm E}^{-1}$ to get sufficient statistics. 

From MD simulation with Yukawa melts the following information was collected: We calculated the RDF $g(r)$ and the normalized velocity autocorrelation function (VAF)
\begin{equation}\label{VAF}
Z(t)=\frac{\langle{\bf v}(t){\bf v}(0)\rangle}{\langle{\bf v}(0)^2\rangle},  
\end{equation}
as well as the mean squared displacement (MSD),  ${\rm MSD}(t) = \langle \Delta {\bf r}^2 (t)\rangle$. Here ${\bf v}(t)$ is the velocity of a given particle at a time $t$, $\Delta {\bf r}={\bf r}(t)-{\bf r}(0)$ is its displacement from the initial position, and $\langle\cdot\cdot\cdot \rangle$ denotes an average over all particles. The time-dependent diffusion coefficients can be evaluated employing either MSD: $D(t)={\rm MSD}(t)/6t$, or VAF: $D(t)=v_{\rm T}^2\int_0^tZ(t')dt'$ (Green-Kubo relation). These definitions are not equivalent, but in the limit $t\rightarrow \infty$ they should approach one single value, the conventional self-diffusion constant $D$.       

\begin{figure}
\includegraphics[width=8cm]{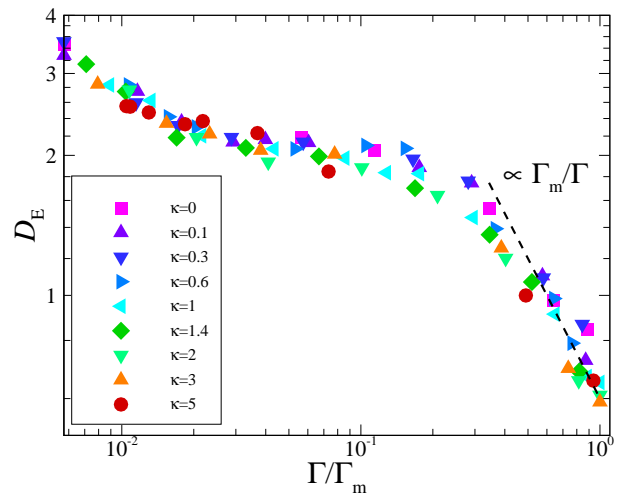}
\caption{Reduced self-diffusion coefficient of Yukawa fluids, expressed in units of $v_{\rm T}^2/\Omega_{\rm E}$, as a function of the relative coupling strength $\Gamma/\Gamma_{\rm m}$. The data points for $\kappa>0$ and $\Gamma\lesssim \Gamma_{\rm m}$ are from Ref.~\onlinecite{OhtaPoP2000}. The data points for $\kappa=0$ (OCP limit) are from Ref.~\onlinecite{HansenPRA1975}. The data points for Yukawa melts ($\Gamma=\Gamma_{\rm m}$) are obtained in this work. The dotted line marks the scaling $D_{\rm E}\propto (\Gamma_{\rm m}/\Gamma)$ in the vicinity of the fluid-solid phase transition. Three regime of self-diffusion can be identified, see the text for details.}
\label{Fig1}
\end{figure}

\section{Results and discussion}\label{Results}

The reduced diffusion coefficient, $D_{\rm E}$, as a function of the reduced coupling parameter $\Gamma/\Gamma_{\rm m}=T_{\rm m}/T$ is shown in Fig.~\ref{Fig1}.  The data points for Yukawa and OCP fluids are taken from Ref.~\onlinecite{OhtaPoP2000} and Ref.~\onlinecite{HansenPRA1975}, respectively. The data points for Yukawa melts ($\Gamma=\Gamma_{\rm m}$) are those obtained in the present MD simulation. The following main trends can be summarized as follows. 

First, we observe that the data points for the dependence of $D_{\rm E}$ on  $\Gamma/\Gamma_{\rm m}$  are grouping around a single universal curve. No systematic dependence on $\kappa$ (that is on the interaction potential softness) is observed.  This should be expected in view of the isomorph theory of systems with hidden scale invariance.~\cite{GnanJCP2009,DyreJPCB2014} Isomorphs
are curves along which certain structural and dynamical properties are approximately invariant in properly reduced units. The melting line is itself an isomorph and thus the lines with the fixed ratios $\Gamma/\Gamma_{\rm m}$ are approximate isomorphs, too. Application of the isomorph theory to Yukawa systems has been recently discussed in detail.~\cite{VeldhorstPoP2015}  

Second, we see that the quantity $D_{\rm E}$ decreases from $\simeq 3$ to $\simeq 0.6$ upon increase in coupling from $0.01\Gamma_{\rm m}$ to $\Gamma_{\rm m}$. This should be compared with almost three orders of magnitude variation of the quantities $D/\omega_{\rm E}a^2$ and $D/\omega_{p}a^2$,~\cite{OhtaPoP2000,DaligaultPRE2012} and  thirty times variation in  $D_{\rm R}$ in a similar range of $\Gamma$.~\cite{RosenfeldPRE2000} Thus, we observe a rather weak dependence of $D_{\rm E}$ on the relative coupling strength $\Gamma/\Gamma_{\rm m}$. This suggests a very rough estimate for the strongly coupled fluid regime,
\begin{equation} \label{estimate}
D \simeq v_T^2/\Omega_{\rm E}, 
\end{equation}
which is accurate to within a factor of two in an extended region of the phase diagram.      

\begin{figure}
\centering
\includegraphics[width=8cm]{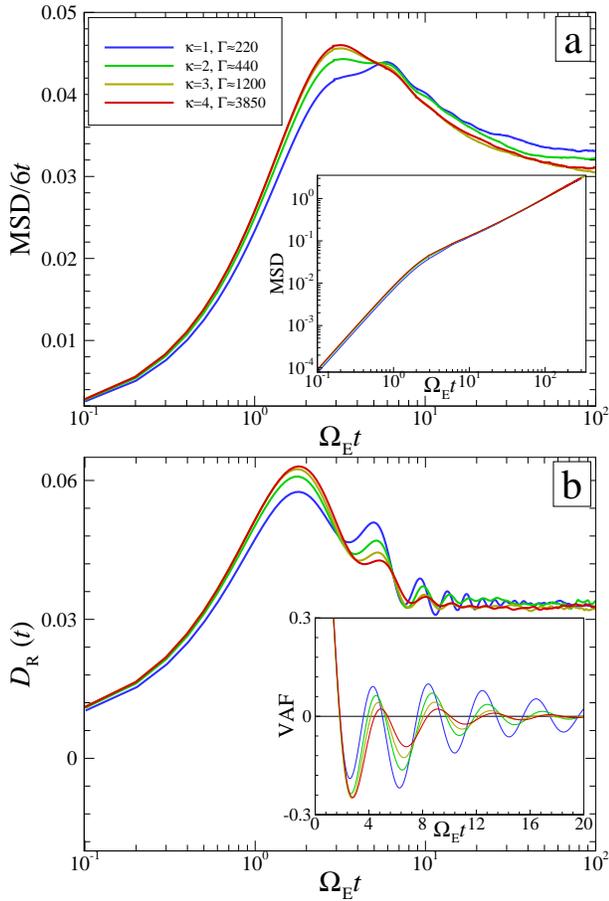}
\caption{Time-dependent reduced self-diffusion coefficient versus the reduced time $t\Omega_{\rm E}$ for Yukawa melts characterized by four different $\kappa$-values: $\kappa = 1$ (blue), $\kappa = 2$ (green), $\kappa = 3$ (gold), and $\kappa = 4$ (red). In (a) the diffusion coefficient is determined from $D(t)={\rm MSD}(t)/6t$, the inset shows reduced ${\rm MSDs}$  as functions of the reduced time in log-log scale. In (b) the diffusion coefficient is determined from $D(t)=v_{\rm T}^2\int_0^tZ(t')dt'$, the inset shows the corresponding velocity autocorrelation functions, $Z(t)$. In both cases the diffusion coefficient is expressed using the Rosenfeld normalization, $D_{\rm R}=D n^{1/3}/v_T$. }
\label{Fig2}
\end{figure}

\begin{figure}
\includegraphics[width=8cm]{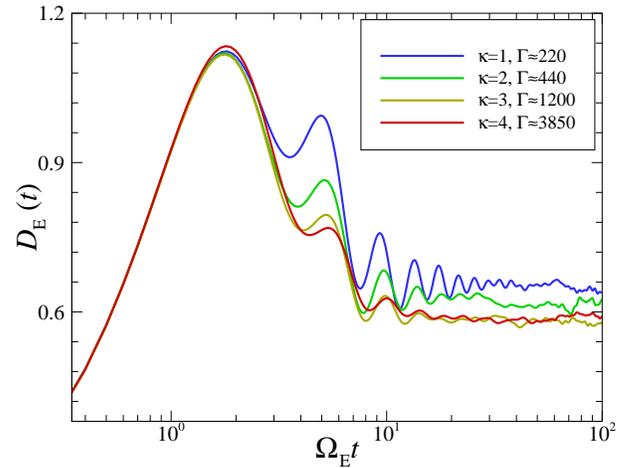}
\caption{Time-dependent reduced self-diffusion coefficient $D_{\rm E}(t)$ versus the reduced time $t\Omega_{\rm E}$ for Yukawa melts characterized by four different $\kappa$-values. The color notation is the same as in Fig.~\ref{Fig2}. The diffusion coefficient is determined from the Green-Kubo relation (i.e. using VAF) and is normalized as $D_{\rm E}=D(\Omega_{\rm E}/v_T^2)$. }
\label{Fig3}
\end{figure}

Third, more careful examination of the data plotted in Fig.~\ref{Fig1} identifies three regimes of self-diffusion. In the weakly coupled regime, $\Gamma\lesssim 0.03\Gamma_{\rm m}$ the diffusion coefficient slowly decreases with coupling. This is the regime where the binary collision approximation for a gaseous-like phase should be appropriate (we did not test this quantitatively, however). Then we observe a plateau-like behavior in a wide regime of coupling strength $0.03\Gamma_{\rm m}\lesssim \Gamma \lesssim 0.3\Gamma_{\rm m}$, corresponding to the fluid phase, where the reduced diffusion constant is almost constant, $D_{\rm E}\simeq 2$. Finally, on approaching the fluid-solid phase transition the reduced diffusion constant roughly scales as $D_{\rm E}\propto (\Gamma_{\rm m}/\Gamma)$ and reaches the value of approximately $D_{\rm E}\simeq 0.6$ at freezing ($\Gamma=\Gamma_{\rm m})$.        

In Figures \ref{Fig2} and \ref{Fig3} we show the results related to Yukawa melts obtained in this work. In figures~\ref{Fig2} time-dependent diffusion coefficients calculated either via MSD (a) or Green-Kubo relation (b) are plotted. The diffusion coefficients are reduced according to the Rosenfeld prescription, $D_{\rm R}=Dn^{1/3}/v_T$.  It is observed that they tend to $D_{\rm R}\simeq 0.03$ as $t\rightarrow \infty$, similar to several other model systems  near the fluid-solid phase transition studied recently (Hertzian, Gaussian-core, and inverse-power-law interactions).~\cite{PondSM2011} The convergence is somewhat better when using Green-Kubo formula [see figure~\ref{Fig2} (b)]. Interestingly, the time-dependent self-diffusion coefficient evaluated at a time point at which the VAF crosses zero [at this point $D(t)$ reaches its first maximum] is roughly twice the actual diffusion constant. Figure~\ref{Fig3} shows essentially the same information as Fig.~\ref{Fig2} (b), but with a different normalization,  $D_{\rm E}$ from Eq.~(\ref{Norm}).  In this case the self-diffusion coefficients approach $D_{\rm E}\simeq 0.6$ as $t\rightarrow \infty$. Somewhat more scattering in $D_{\rm E}$ values is observed compared to $D_{\rm R}$ [compare Fig.~\ref{Fig2} (b) and Fig.~\ref{Fig3}]. The oscillations in time-dependent diffusion coefficients are related to the oscillations in the velocity autocorrelation function, which are characteristic for the caged short-time dynamics of particles in the fluid state.            

The values of differently normalized self-diffusion constants at freezing can be related using the following simple empirical consideration. Near the melting point of a crystalline lattice, an average deviation of particles from their lattice sites is roughly related to the temperature via 
\begin{equation}
\frac{1}{2}m\Omega_{\rm E}^2\langle \delta r^2\rangle \sim \frac{3}{2}T.
\end{equation}
The Einstein frequencies of a dense fluid near freezing and of a crystalline solid near melting (at the melting temperature) are close. According to the Lindemann melting rule we expect at melting $\langle \delta r^2\rangle \sim L^2\Delta^2$, where $L\sim 0.1$ is the Lindemann parameter and $\Delta=n^{-1/3}$ is the mean interparticle separation. Using this to estimate the Einstein frequency at the melting temperature we obtain
\begin{equation}
D\simeq 0.6\frac{v_T^2}{\Omega_{\rm E}}\simeq \frac{0.6L}{\sqrt{3}}v_T\Delta \simeq 0.03 v_Tn^{-1/3}.
\end{equation}
This is exactly what we have observed for Yukawa melts. Since Yukawa interaction has not been assumed in this consideration, neither explicitly, nor implicitly it would make sense to test whether the derived relations above can operate in real systems as well. We have estimated the values of the reduced self-diffusion coefficient of some liquid metals at the freezing point using the data summarized by March and Tosi~\cite{MarchBook} (see Table 5.2 of their book).  These results are shown in Table~\ref{Tab1}. Except the two extreme cases of Cu ($D_{\rm R}\simeq 0.040$) and Zn ($D_{\rm R}\simeq 0.027$) all the other data are scattered in a rather narrow range ($D_{\rm R}\simeq 0.032 \pm 0.03$). Moreover, using a more recent value for the self-diffusion in Cu melt ($3.27\times 10^{-5}$ cm$^2$s$^{-1}$ according to Ref.~\onlinecite{Meyer2015}) we obtain $D_{\rm R}\simeq 0.033$, which also belongs to the specified range.  Similar relationships for the reduced coefficient of self-diffusion in liquid metals at the melting temperature were proposed in Refs.~\onlinecite{AscarelliPR1968,MarchJCP1984} on the basis of quite different arguments.  In case of extremely soft Coulomb interaction, corresponding to the OCP limit, the self-diffusion coefficient at the fluid-solid phase transition is $D_{\rm R}\simeq 0.031$. This value has been obtained using the fitting formula by Daligault,~\cite{DaligaultPRE2012} adopting $\Gamma_{\rm m}\simeq 172$ at the fluid-solid phase transition.~\cite{HamaguchiPRE1997} The hard-sphere result for the self-diffusion near freezing is somewhat smaller $D_{\rm R}\simeq 0.02$,~\cite{HeyesJPCB2007} possibly signaling the crossover between soft sphere and hard sphere regimes of collective motion and transport.~\cite{KhrapakSciRep2017}   All this implies that certain tendencies identified in this work are unlikely specific to the Yukawa interaction potential, but can be applicable to other model and real systems, provided the interactions are soft enough. 

\begin{table}
\caption{\label{Tab1} Reduced diffusion constants $D_{\rm R}$ of ten liquid metals at the corresponding melting points as calculated from the data summarized in Ref.~\onlinecite{MarchBook}. }
\begin{ruledtabular}
\begin{tabular}{cccccccccc}
Li & Na & K & Rb & Cu & Ag & Pb & Zn & In & Hg \\ \hline
0.029 & 0.033 & 0.032 & 0.034 & 0.040 & 0.031 & 0.035 & 0.027 & 0.032 & 0.034 \\
\end{tabular}
\end{ruledtabular}
\end{table}

To conclude the discussion, we point out one more potential application of the obtained results. The discovered dependence of $D_{\rm E}$ on relative coupling strength, combined with the elementary expression for the diffusion coefficient $D=T/m \xi$, allows us to estimate the characteristic momentum transfer frequency in Yukawa fluids, 
\begin{displaymath}
\xi = \Omega_{\rm E}/D_{\rm E}.      
\end{displaymath}
We get in the simplest rough approximation $\xi \sim  \Omega_{\rm E}$, improvements can be made by making use of the data plotted in Fig.~\ref{Fig1}. This can be relevant for the momentum transfer frequency in particle-particle collisions in complex (dusty) plasmas. Previously, a unified theory of momentum transfer in complex plasmas has been developed using the binary collision approach, assuming the interaction potential between charged species of the Yukawa type.~\cite{KhrapakPRE2004,MorfillPS2004,KhrapakAIP2005,NRL} This is normally adequate for electron-particle and ion-particle collisions, but the binary collision approximation for particle-particle collisions can only be relevant for extremely rarefied particle clouds with weak electrostatic coupling between them. The results presented in this article suggest a useful method to evaluate momentum transfer rates beyond the weak coupling regime.        

\section{Conclusion}\label{Concl}

To summarize, we have tested de Gennes approximation for the self-diffusion coefficient in simple fluids [Eq.~(\ref{Diff_DG})] using the data for self-diffusion in one-component Yukawa systems. Equation (\ref{Diff_DG}) is shown to be inexact, but nevertheless useful. It turns out to be accurate to within a factor of two in the entire strongly coupled fluid regime. 

From the simulations of Yukawa melts we have observed that the properly normalized self-diffusion coefficient is approximately constant. The value $D_{\rm R}\simeq 0.03$ is consistent with the values reported for other simple model systems  at freezing, such as Hertzian, Gaussian-core, and inverse-power-law fluids.~\cite{PondSM2011} It is also consistent with the self-diffusion coefficient measured in liquid metals at the melting temperature, giving strong support to the emerging dynamic freezing criterion for simple atomistic systems. This can potentially complement the well-known empirical dynamic freezing criterion for over-damped colloidal fluids.~\cite{LowenPRL1993} Simple explanation for the observed universality of the self-diffusion coefficient has been suggested.     

Finally, it has been pointed out that the obtained results can be of some use in estimating the momentum transfer rates in particle-particle collisions in complex (dusty) plasmas. The existing approaches have been often limited to the binary collision approximation, only appropriate for the weakly coupled gaseous regime. The observed trends in self-diffusion can be useful in estimating momentum transfer rates in the strongly coupled fluid regime. The reported results can also be of some use in the context of self-diffusion in  strongly coupled plasmas, experimental measurements of which have been recently reported.~\cite{StricklerPRX2016}

\begin{acknowledgments}
This work was supported by the A*MIDEX project (Nr.~ANR-11-IDEX-0001-02) funded by the French Government ``Investissements d'Avenir'' program managed by the French National Research Agency (ANR). Structure and dynamical data analysis was supported by the  Russian Science Foundation, Project RSF 14-50-00124. We thank Roman Kompaneets for reading the manuscript and suggesting improvements. 
\end{acknowledgments}

\bibliographystyle{aipnum4-1}
\bibliography{SD_Ref}

\end{document}